\documentclass[12pt,letterpaper]{article}
\pdfoutput=1
\usepackage{jheppub}
\usepackage[latin1]{inputenc}
\usepackage{bbm,amsfonts}
\usepackage{graphicx}
\usepackage{amssymb,amsmath}
\usepackage{fancybox}

\author[a]{Marco S. Bianchi,}
\author[b]{ Luca Griguolo,}
\author[c]{ Matias Leoni,}
\author[d]{Andrea Mauri,}
\author[d,e]{\\Silvia Penati}
\author[f]{and Domenico Seminara}
 \preprint{QMUL-PH-16-07}
\affiliation[a]{Center for Research in String Theory - School of Physics and Astronomy Queen Mary University of London, Mile End Road, London E1 4NS, UK}
\affiliation[b]{Dipartimento di Fisica e Scienze della Terra, Universit\`a di Parma and INFN Gruppo Collegato di Parma, Viale G.P. Usberti 7/A, 43100 Parma, Italy}
\affiliation[c]{Physics Department, FCEyN-UBA \& IFIBA-CONICET\ 
  Ciudad Universitaria, Pabell\'on I, 1428, Buenos Aires, Argentina }
\affiliation[d]{ Dipartimento di Fisica, Universit\`a degli studi di Milano--Bicocca, Piazza della Scienza 3, I-20126 Milano, Italy }
\affiliation[e]{INFN, Sezione di Milano--Bicocca, Piazza della Scienza 3, I-20126 Milano, Italy }
\affiliation[f]{Dipartimento di Fisica, Universit\`a di Firenze and INFN Sezione di Firenze, via G. Sansone 1, 50019 Sesto Fiorentino, Italy\\}

\emailAdd{m.s.bianchi@qmul.ac.uk}  
\emailAdd{luca.griguolo@pr.infn.it} 
\emailAdd{andrea.mauri@mi.infn.it} 
\emailAdd{leoni@df.uba.ar}
\emailAdd{silvia.penati@mib.infn.it} 
\emailAdd{seminara@fi.infn.it}

\abstract{Supersymmetric localization provides exact results that should match QFT computations in some regularization scheme. The agreement is particularly subtle in three dimensions where complex answers from localization procedure sometimes arise. We investigate this problem by studying the expectation value of the 1/6 BPS Wilson loop in planar ABJ(M) theory at three loops in perturbation theory. We reproduce the corresponding term in the localization result and argue that it originates entirely from a non--trivial framing of the circular contour. Contrary to pure Chern-Simons theory, we point out that for ABJ(M) the framing phase is a non--trivial function of the couplings and that it potentially receives contributions from vertex-like diagrams. Finally, we briefly discuss the intimate link between the exact framing factor and the Bremsstrahlung function of the 1/2-BPS cusp.}

\title{Framing and localization in Chern-Simons theories with matter}

\keywords{Chern--Simons matter theories, BPS Wilson loops, framing, localization}

\csname @addtoreset\endcsname{equation}{section}

\def\bseq{\begin{subequation}}  
\def\eseq{\end{subequation}}
\def\bsea{\begin{subeqnarray}}  
\def\esea{\end{subeqnarray}}

\newcommand{\beq}{\begin{equation}}
\newcommand{\bea}{\begin{eqnarray}}
\newcommand{\eea}{\end{eqnarray}}
\newcommand{\eeq}{\end{equation}}

\newcommand {\non}{\nonumber}

\renewcommand{\a}{\alpha}
\renewcommand{\b}{\beta}

\renewcommand{\d}{\delta}
\newcommand{\pa}{\partial}
\newcommand{\g}{\gamma}
\newcommand{\G}{\Gamma}

\newcommand{\e}{\epsilon}

\renewcommand{\l}{\lambda}

\newcommand{\m}{\mu}

\newcommand{\n}{\nu}

\newcommand{\s}{\sigma}

\def\Mb{\kern 2pt\mathchoice
        {
         \vbox{\hrule width10pt height 0.4pt depth 0pt
         \kern 1.2pt\hbox{\kern -2pt$\displaystyle M$}}}
        {
         \vbox{\hrule width10pt height 0.4pt depth 0pt
         \kern 1.2pt\hbox{\kern -2pt$\textstyle M$}}}
        {
\vbox{\hrule width6pt height 0.4pt depth 0pt
         \kern 1.0pt\hbox{\kern -2pt$\scriptstyle M$}}}
        {
         \vbox{\hrule width5pt height 0.4pt depth 0pt
         \kern 0.8pt\hbox{\kern -2pt$\scriptscriptstyle M$}}}}

\def\Sb{\kern 2pt\mathchoice
        {
         \vbox{\hrule width6pt height 0.4pt depth 0pt
         \kern 1.2pt\hbox{\kern -2pt$\displaystyle S$}}}
        {
         \vbox{\hrule width6pt height 0.4pt depth 0pt
         \kern 1.2pt\hbox{\kern -2pt$\textstyle S$}}}
        {
         \vbox{\hrule width3.5pt height 0.4pt depth 0pt
         \kern 1.0pt\hbox{\kern -2pt$\scriptstyle S$}}}
        {
         \vbox{\hrule width3pt height 0.4pt depth 0pt
         \kern 0.8pt\hbox{\kern -2pt$\scriptscriptstyle S$}}}}

\def\Rb{\kern 2pt\mathchoice
        {
         \vbox{\hrule width5.5pt height 0.4pt depth 0pt
         \kern 1.2pt\hbox{\kern -2.5pt$\displaystyle R$}}}
        {
         \vbox{\hrule width5.5pt height 0.4pt depth 0pt
         \kern 1.2pt\hbox{\kern -2.5pt$\textstyle R$}}}
        {
         \vbox{\hrule width3.5pt height 0.4pt depth 0pt
         \kern 1.0pt\hbox{\kern -2.2pt$\scriptstyle R$}}}
        {
         \vbox{\hrule width3pt height 0.4pt depth 0pt
         \kern 0.8pt\hbox{\kern -2.2pt$\scriptscriptstyle R$}}}}

  \def\pp{{\mathchoice
      {
          \kern 1pt%
          \raise 1pt
          \vbox{\hrule width5pt height0.4pt depth0pt
            \kern -2pt
            \hbox{\kern 2.3pt
              \vrule width0.4pt height6pt depth0pt
              }
            \kern -2pt
            \hrule width5pt height0.4pt depth0pt}%
            \kern 1pt
       }
        {
          \kern 1pt%
          \raise 1pt
          \vbox{\hrule width4.3pt height0.4pt depth0pt
            \kern -1.8pt
            \hbox{\kern 1.95pt
              \vrule width0.4pt height5.4pt depth0pt
              }
            \kern -1.8pt
            \hrule width4.3pt height0.4pt depth0pt}%
            \kern 1pt
        }
        {
          \kern 0.5pt%
          \raise 1pt
          \vbox{\hrule width4.0pt height0.3pt depth0pt
            \kern -1.9pt  
            \hbox{\kern 1.85pt
              \vrule width0.3pt height5.7pt depth0pt
              }
            \kern -1.9pt
            \hrule width4.0pt height0.3pt depth0pt}%
            \kern 0.5pt
        }
        {
          \kern 0.5pt%
          \raise 1pt
          \vbox{\hrule width3.6pt height0.3pt depth0pt
            \kern -1.5pt
            \hbox{\kern 1.65pt
              \vrule width0.3pt height4.5pt depth0pt
              }
            \kern -1.5pt
            \hrule width3.6pt height0.3pt depth0pt}%
            \kern 0.5pt
        }
    }}

  \def\mm{{\mathchoice
               {
                 \kern 1pt
           \raise 1pt    \vbox{\hrule width5pt height0.4pt depth0pt
                  \kern 2pt
                  \hrule width5pt height0.4pt depth0pt}
                 \kern 1pt}
               {
                \kern 1pt
           \raise 1pt \vbox{\hrule width4.3pt height0.4pt depth0pt
                  \kern 1.8pt
                  \hrule width4.3pt height0.4pt depth0pt}
                 \kern 1pt}
               {
                \kern 0.5pt
           \raise 1pt
                \vbox{\hrule width4.0pt height0.3pt depth0pt
                  \kern 1.9pt
                  \hrule width4.0pt height0.3pt depth0pt}
                \kern 1pt}
               {
               \kern 0.5pt
         \raise 1pt  \vbox{\hrule width3.6pt height0.3pt depth0pt
                  \kern 1.5pt
                  \hrule width3.6pt height0.3pt depth0pt}
               \kern 0.5pt}
               }}

\def\pd{{\kern0.5pt
           + \kern-5.05pt \raise5.8pt\hbox{$\textstyle.$}\kern
0.5pt}}

\def\pmd{{\kern0.5pt
          \pm \kern-5.05pt
\raise6.3pt\hbox{$\textstyle.$}\kern1.5pt}}

\def\md{{\mathchoice
   {
      {{\kern 1pt - \kern-6.2pt \raise5pt\hbox{$\textstyle.$}\kern
1pt}}}
    {
      {{\kern 1pt - \kern-6.2pt \raise5pt\hbox{$\textstyle.$}\kern
1pt}}}
    {
      {\kern0.5pt - \kern-5.05pt
\raise3.4pt\hbox{$\textstyle.$}\kern0.5pt}}
    {
      {\kern0.5pt - \kern-5.05pt
\raise3.4pt\hbox{$\textstyle.$}\kern0.5pt}}}}

\def\beq{\begin{equation}}
\def\eeq{\end{equation}}
\def\bea{\begin{eqnarray}}
\def\eea{\end{eqnarray}}
\def\Tr{\textstyle{\rm Tr}}

\def\a{\alpha}
\def\b{\beta}
\def\g{\gamma}

\def\d{\delta}
\def\e{\epsilon}

\def\l{\lambda}
\def\G{\Gamma}

\begin{document}
\maketitle

\section{Introduction and Conclusions}

In three dimensional Chern-Simons theories framing factors usually appear in the evaluation of Wilson Loop operators (WL) on non--intersecting curves, being them associated to regularization ambiguities in the contour integrations. 

This has been extensively studied for pure topological Chern--Simons (CS) theories, for which the first evidence of framing goes back to the seminal paper by Witten \cite{Witten}. The exact expression for the vacuum expectation value $\langle \mathcal{W} \rangle$ obtained by using non--perturbative methods contains in fact an overall phase factor which is not topologically invariant, being induced by the gauge fixing procedure that necessarily introduces a metric dependence. This factor can be made topologically invariant by framing the original manifold.  

From a quantum field theory point of view the appearance of these factors has a very clear explanation \cite{Witten, GMM, Labastida}. Correlation functions of gauge connections $\langle A_{\mu_1}(x_1) \cdots A_{\mu_n}(x_n) \rangle$ entering the perturbative expansion of  a WL require a regularization prescription in order to be well--defined at coincident points on the contour. 
One possibility is to use point--splitting regularization that allows each gauge connection to run on a deformed contour (frame), slightly displaced and possibly intertwined with the original one. When the regularization is removed, framing--dependent but metric--independent terms survive that are expressible as powers of the linking number, that is the number of windings of the deformed path around the original one. 
It has been proved \cite{GMM, Labastida} that, resumming the perturbative series, these terms exponentiate the one--loop contribution.

The fact that in the pure CS case the framing factor turns out to be a one--loop effect relies on two important properties of the perturbative series: 1) Framing--dependent contributions come only from diagrams containing collapsible propagators \cite{GMM, Labastida}, that is propagators joining two points on the contour that can get together; 2)
In Landau gauge where these calculations are performed, the gauge propagator is one--loop exact (in any regularization scheme that preserves scale and BRS invariance it does not acquire any infinite or finite quantum corrections \cite{Blasi:1989mw, Delduc:1990je, Chen:1992ee}, otherwise it may acquire a finite, scheme--dependent one--loop quantum correction \cite{Guadagnini:1989kr, Chen:1992ee, Labastida}). 

This pattern is no longer true in CS theories coupled to interacting matter.
In fact, when matter is present non-trivial higher--loop corrections to the vector propagator generally appear. This is for instance the case in ${\cal N}=2,3$ $U(N)$ CS theories \cite{Gaiotto-Yin}, ${\cal N}=4$ quiver CS--matter theories \cite{Hosomichi:2008jd} and ${\cal N}=6$ ABJ(M) theories \cite{ABJM, ABJ}. Moreover, matter interaction vertices give rise to new topologies of diagrams that in principle might be framing dependent, although not containing collapsible gauge propagators. 

It is then natural to ask how the framing dependence in WL gets modified in the presence of matter and whether it can still be factorized as a phase possibly given in terms of a quantum corrected framing function.  

In this paper we are  going to investigate this problem by studying  the bosonic 1/6 BPS Wilson loop in ABJ(M) theory \cite{Drukker, Chen, Rey} in the planar limit.  Using dimensional regularization with dimensional reduction (DRED) we perform a three--loop calculation, as this is the first non--trivial order where framing due to matter may arise. Moreover, since an exact result for the 1/6 BPS WL is available from localization, comparing our genuine perturbative calculation with the weak coupling expansion of the matrix model allows to identify the framing contributions in the localization result.

First of all, we compute the two--loop correction to the gauge propagator. Although most of the contributing integrals are UV divergent, in DRED scheme their sum turns out to be finite and non--vanishing. This result is then used to evaluate the diagram contributing to $\langle \mathcal{W} \rangle$ at third order given by the exchange of a (collapsible) two--loop effective propagator. Two classes of framing dependent contributions arise, proportional to $\l_1 \l_2^2$ and $\l_1^2 \l_2$ respectively, where $\l_{1,2}$ are the `t Hooft couplings of  ABJ. Using framing regularization for splitting contours, once the framing parameter is removed the result ends up being proportional to the the linking number (Gauss integral in eq. (\ref{gauss})). 

Comparing with the third order expansion of the matrix model result \cite{Marino:2009jd,Drukker:2010nc} we find that the perturbative contribution proportional to the color factor $\l_1 \l_2^2$ reproduces exactly the localization result, once we choose the linking number to be (minus) one. This matching not only represents a non--trivial check of the matrix model calculation at three loops, but it also allows to identify the imaginary contribution appearing at third order in the weak coupling expansion of the matrix model as a genuine framing contribution. Moreover, it confirms that DRED scheme is consistent with localization, as already found at lower loops \cite{Bianchi:2013zda, BGLP2, GMPS}. 

The factorization theorems \cite{Labastida} that in the pure CS case were at the basis of the exponentiation of the one--loop framing contribution, are still at work in the presence of finite quantum corrections to the gauge propagator. Therefore, the third order contribution to $\langle \mathcal{W} \rangle$ is the first non--trivial term in the expansion of an exponential that corrects the original one--loop framing coming from the pure CS sector.  The result in fundamental representation can then be written as    
\beq
\langle \mathcal{W} \rangle = e^{i \pi \left( \l_1 - \frac{\pi^2}{2} \l_1\l_2^2 + {\cal O}(\l^5) \right) \chi(\G, \G_f)} \left( 1 - \frac{\pi^2}{6} (\l_1^2 - 6 \l_1\l_2)  + {\cal O}(\l^4) \right),
\eeq
where $\chi(\G, \G_f)$ is the linking number between the original and the framing contour.

However, this is not the end of the story. As already mentioned, the perturbative result at three loops contains an extra framing--dependent term proportional to $\l_1^2 \l_2$ that does not have a counterpart in the weak coupling expansion of the matrix model. Therefore, there must be some other contribution at this order that cancels the extra term in the perturbative result. Having exhausted the topologies with collapsible propagators, the only possibility is that non--trivial contributions to framing arise also from vertex--like diagrams. A complete analytical calculation of all the contributions and the check of the actual cancellation of framing in this sector is out of the scope of this paper. 
However, we perform a numerical investigation of possible vertex--like diagrams contributing in this sector and, in fact, we find that the corresponding integrals depend linearly on the linking number. This evades the theorem of \cite{Labastida} that is valid in the pure CS case and represents a novel feature of CS theories with matter that deserves further investigation. 

Furthermore, we notice an interesting relation between our results and a recent proposal for the exact Bremsstrahlung function  $B_{1/2}(\lambda)$ in ABJM theory \cite{Bianchi:2014laa}.
There, a general formula for $B_{1/2}(\lambda)$ encoding the near-BPS limit of the cusp anomalous dimension for fermionic Wilson Loop operators,
has been derived in analogy with the ${\cal N}=4$ SYM \cite{Correa:2012at}. The explicit construction of ``latitude" fermionic 1/6 BPS Wilson Loops \cite{Cardinali:2012ru} was
taken into account together with some reasonable assumptions on their near-maximal circle behavior (see \cite{Bianchi:2014laa} details). The final answer,
consistent with two-loop Feynman diagrams computations \cite{Griguolo:2012iq} and leading \cite{Forini:2012bb} and subleading \cite{Aguilera-Damia:2014bqa} strong coupling expansions reads
\begin{equation}
\label{Bfin}
B_{1/2}(\lambda)= \frac{1}{8\pi} \, {\rm tg}\Phi_B
\end{equation}
where the Bremsstrahlung function is completely expressed in terms of the phase $\Phi_B$ of the 1/6 BPS bosonic loop on the maximal circle.
According to the results presented in this paper it is reasonable to expect that the whole phase be a framing effect, as explicitly seen at
three-loop order. On the other hand, in the near-BPS cusp computation on the plane, framing regularization appears to play no particular role
while fermionic interactions, absent in 1/6 BPS bosonic case here, are essential to recover the result. A possible relation between these two
apparently unrelated contributions seems therefore suggested and certainly deserves a deeper analysis. It would be interesting to explore if a
similar relation emerges for the Bremsstrahlung function at generic opening angle \cite{Correa:2012at}, investigating wedge 1/6 BPS fermionic Wilson loops
on $S^2$. In deriving eq. (\ref{Bfin}) it was also used the explicit vanishing of certain derivatives of $n$-winding 1/6 BPS bosonic Wilson loops: again this
vanishing crucially depends on the framing nature of some contributions \cite{Klemm:2012ii}. A closer inspection of framing effects for $n$-winding BPS Wilson
Loops in ABJ(M) is  therefore certainly worthwhile \cite{MB}. 
 
We conclude by observing that the present analysis can be extended to other interesting cases, most notably the 1/2 BPS Wilson loop in ABJ(M). The two-loop matching with the localization result has been carefully discussed in \cite{Bianchi:2013zda,BGLP2,GMPS} at framing zero. It would be interesting to make an explicit diagramatic check at non--trivial framing. New supersymmetric Wilson loops in ${\cal N}=4$ super Chern-Simons \cite{OWZ1, Cooke:2015ila, OWZ2, Griguolo:2015swa, OWZ3} would also reserve surprises at three-loop \cite{New}.

\section{Framing factors in Wilson loops}
 
As extensively discussed in literature \cite{Witten, GMM, Labastida}, in pure Chern--Simons theory the vacuum expectation value of Wilson loop operators on close paths $\G$
\bea
\label{WLpure}
\langle \mathcal{W_{\rm CS}} \rangle &=& \frac{1}{N} \int [{\cal D} A] \; e^{-S_{CS}}  \; \Tr  \, P \exp{\left( - i \int_\G dx^\mu A_\mu(x)  \right) } 
\eea
is affected by finite regularization ambiguities if point--splitting is used to regularize short distance singularities in  $\langle A_{\mu_1}(x_1) \cdots A_{\mu_n}(x_n) \rangle$ which could potentially appear when multiple points on $\G$ clash. In this regularization scheme this is avoided by requiring every single point $x_i$ to run on a different path (called frame). For instance, in the first non--trivial  correction  proportional to the tree--level propagator $\langle A_{\mu_1}(x_1) A_{\mu_2}(x_2)\rangle$ the second gauge connection can be chosen running on an infinitesimal deformation of the original path defined by \cite{Witten, GMM}
\beq
\label{frame2}
\G_f : \quad x^\mu(\tau) \rightarrow y^\mu(\tau) = x^\mu(\tau) + \a \, n^\mu(\tau) \quad , \quad |n(\tau)| = 1
\eeq
where $n^\mu(\tau)$ is orthogonal to the path $\G$. 
 
Although in pure CS no divergences appear \cite{GMM}, the removal of the point-splitting regularization ($\a \to 0$) at the end of the calculation leaves a deformation--dependent term which is proportional to the linking number of the two non-intersecting closed paths, the original $\G$ and the deformed $\G_f$. This is given by the Gauss integral
\beq
\label{gauss}
\chi(\G, \G_f) = \frac{1}{4\pi} \oint_{\G} dx^\m \oint_{\G_f} dy^\n \; \varepsilon_{\m\n\rho} \frac{(x-y)^\rho}{|x-y|^3} 
\eeq
It is a topological invariant that takes integer values corresponding to the number of times the path $\G_f$ winds around $\G$. 

Diagrams associated to higher--loop corrections $\langle A_{\mu_1}(x_1) \cdots A_{\mu_n}(x_n) \rangle$  containing at least one collapsible gauge propagator \footnote{Following Ref. \cite{Labastida} we name ``collapsible propagator" any free propagator that connects two different points on the WL contour which can get together.} lead to frame--dependent terms (see examples in Fig. 1).
The rest of contributions have been argued to be framing independent \cite{GMM, Labastida}. 
The framing dependent terms contain powers of $\chi(\G, \G_f)$ with the right coefficients to be factorized as an overall  phase. Therefore in pure CS theory the framing dependence appears in a very controlled way, as for instance for the $U(N)$ case the exact vacuum expectation value in the fundamental representation takes the form \footnote{Here and in the following $k$ must be understood as the renormalized coupling constant. It coincides with  its bare values $k_{B}$ if we use DRED scheme or with $k_{B}+N$ if we instead employ higher derivative or massive regularization.}
\beq
\label{wit}
\langle \mathcal{W_{\rm CS}} \rangle = e^{\frac{\pi i N }{k} \chi(\G, \G_f)} \,   \rho(\G)
\eeq
where $\rho$ is a frame--independent function.  

This result is supported by two--loop calculations \cite{GMM, Labastida, Chen:1992ee}. An all--loop proof has also been given \cite{Labastida}, which is based on the following general properties of the perturbative series:

\noindent
(1) The gauge propagator does not acquire any quantum correction beyond one-loop (which is for instance true in Landau gauge, using DRED scheme \cite{Blasi:1989mw, Chen:1992ee}, where even the one-loop correction vanishes). 

\noindent
(2)  A diagram gives framing factors if and only if it contains at least one collapsible propagator. 

\noindent
(3) Reducible diagrams containing separated sub--diagrams factorize into the product of partial contributions associated to each sub--diagram. 

In particular, the second statement (very reasonable although not rigorously proved, as far as we know) prevents any vertex--like diagram with no isolated propagators from contributing to framing.  

\begin{figure}[h]
\begin{center}
 \includegraphics[width=0.50\textwidth]{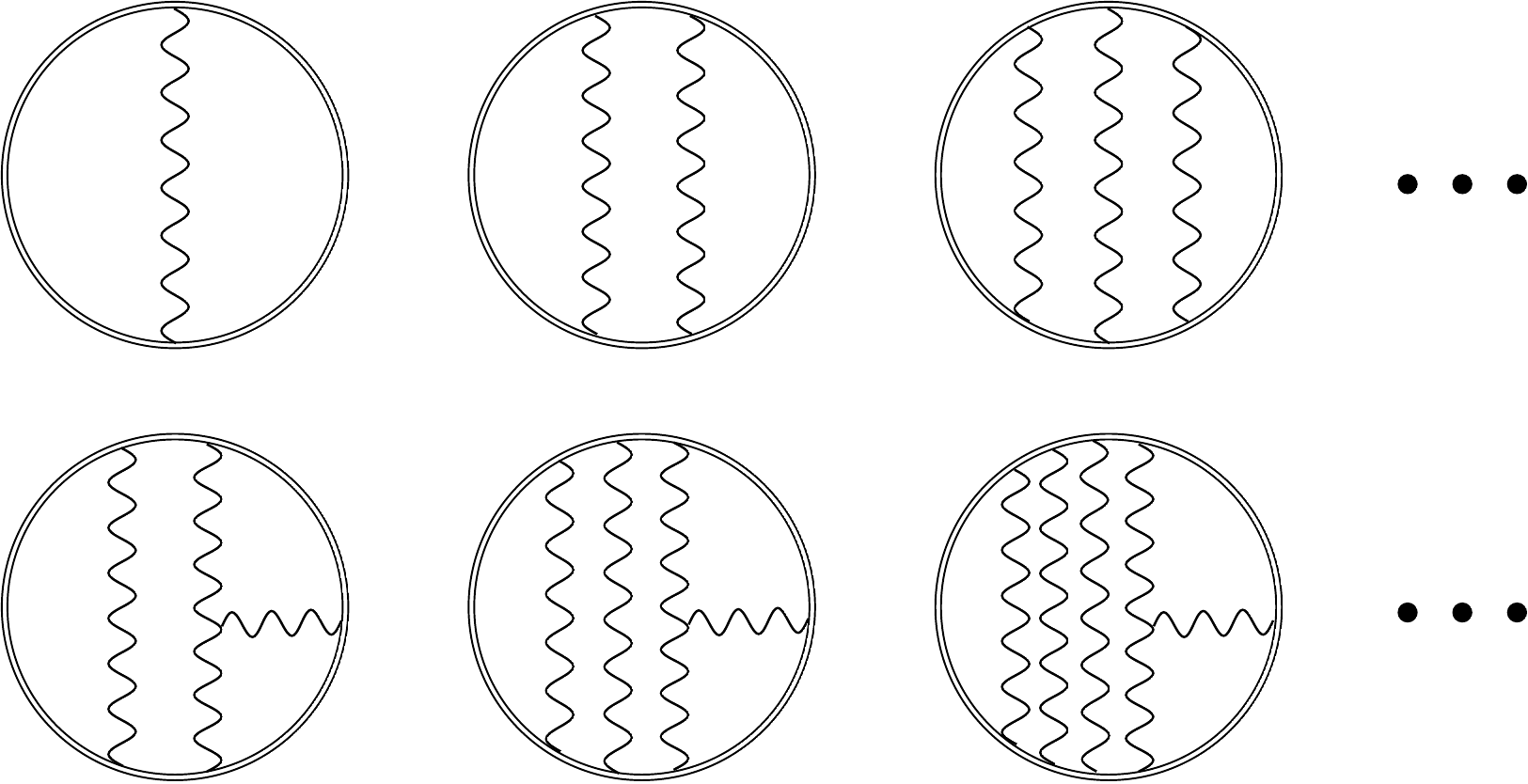}
\caption{Examples of diagram series with collapsible propagators giving framing dependent contributions. Each picture should be meant as indicating the diagram plus all possible permutations of contour points.}\label{expo}
\end{center}
\end{figure}
 
It is important to note that the tensorial structure of the tree--level vector propagator in Landau gauge (see eq. (\ref{treevector})) plays a crucial role in determining the framing factor. It is in fact the $\varepsilon_{\m\n\rho}$ tensor that is responsible for the reconstruction of the linking number in eq. (\ref{gauss}).  

In supersymmetric pure CS theories a similar pattern appears and the identification of the correct framing factor is confirmed by an exact calculation done using localization techniques \cite{Kapustin, Drukker:2010nc, Marino:2011nm}. We recall that the result from localization is necessarily at framing -1, as the only point--splitting regularization compatible with the supersymmetry used to localize the functional integral on $S^3$ is the one where the path and its frame wrap different Hopf fibers \cite{Kapustin}.  

The structure of the framing factor  in susy and non-susy pure CS theories heavily relies on the fact that in Landau gauge these theories are all--loop finite and in dimensional reduction scheme not even finite corrections to the vector propagator seem to arise \cite{Chen:1992ee, Blasi:1989mw} (statement (1) above). In fact, this implies that the $1/k$ effect coming from the exchange of a tree--level propagator, eventually exponentiated by summing all order diagrams as in Fig. \ref{expo}, is the only possible source of framing.  

The situation drastically changes in CS theories with matter where the vector propagator can get finite (or infinite) loop corrections. In this case the vector propagators appearing in the framing dependent diagrams of Fig. \ref{expo} should be replaced by effective propagators, which are power series in $1/k$. Still, we may expect that the factorization of reducible diagrams  works and that the coefficients are the right ones to exponentiate the result from the exchange of a single, effective propagator (statements (2) and (3) above).   As a result, a framing phase of the form $\exp{(i f(1/k) \chi(\G, \G_f))}$ should arise, where the {\em framing function} $f(1/k)$ is a power series in $1/k$ inherited from perturbative corrections to the propagator. 
However, in the presence of interacting matter we are not guaranteed {\em a priori} that perturbative contributions to the framing function only come from diagrams with collapsible propagators (statement (2) above) and novel framing factors could arise. 

In order to investigate these questions, we focus on well--known CS theories with matter, that is ABJ(M) models.  

In $U(N_1)_k \times U(N_2)_{-k}$ ABJ(M) theory, the $U(N_1)$ 1/6-BPS Wilson loop is defined as \cite{Rey, Drukker, Chen}
\bea
\label{WL}
\langle \mathcal{W} \rangle &=& \frac{1}{N_1} \int [{\cal D} (A, \hat{A}, C, \bar{C}, \psi, \bar{\psi})] \; e^{-S}  \; \Tr \left[ P \exp{
\left( - i \int_\G d\tau {\cal A}(\tau) \right) } \right]
\eea
where the euclidean action is given in eq. (\ref{action}) and the generalized connection  
\beq
\label {connection}
{\cal A} =  A_\mu \dot{x}^\mu - \frac{2\pi i}{k}  |\dot{x}| M_J^{\; I} C_I \bar{C}^J 
\eeq
contains non--trivial couplings to the scalar matter fields governed by the matrix $M_I^{\; J} = {\rm diag}(+1,+1,-1,-1)$. 
The path $\G$ is the unit circle parametrized as 
\beq
x^\mu(\tau) = (0, \cos{\tau}, \sin{\tau})
\eeq
Similarly, a second 1/6-BPS Wilson loop can be defined by simply replacing the $U(N_1)$ connection $A_\mu$ with the $U(N_2)$ connection $\hat{A}_\mu$ in eq. (\ref{connection}) and changing the scalar couplings accordingly. However, for the scopes of our discussion we can just focus on the first one. 

The quantity in (\ref{WL}) has been evaluated in DRED scheme and at framing zero, perturbatively up to three loops for the ABJM model \cite{Rey} and more generally for the ABJ one \cite{Bianchi:2013zda, BGLP2, GMPS}. The result reads
\beq
\label{WLpert}
\langle \mathcal{W} \rangle = 1 - \frac{\pi^2}{6} (\l_1^2 - 6 \l_1 \l_2)   + {\cal O}(\l^4)
\eeq
where $\l_1 = N_1/k, \l_2 = N_2/k$.
Moreover, a general analysis based on the counting of  $\varepsilon$ tensors together with the planarity of the contour $\G$ and the identity $\Tr M^{2n+1}=0$ rules out any perturbative contribution at odd loops \cite{Rey}.

Expression (\ref{WL}) has also been evaluated using localization techniques for three--dimensional, ${\cal N} \geq2$ supersymmetric CS--matter theories \cite{Kapustin}.
From the matrix model result of \cite{Drukker:2010nc} expanded at weak coupling one obtains (the expansion at higher orders is given in Appendix \ref{expansion})
\begin{equation}\label{WLMM}
\langle \mathcal{W} \rangle = e^{\pi i \lambda_1} \left( 1-\frac{\pi^2}{6} (\l_1^2 - 6 \l_1 \l_2)  - i \frac{\pi^3 }{2} 
\l_1 \l_2^2 + {\cal O}(\l^{4}) \right)
\end{equation}
where the standard framing--one factor for pure CS has been stripped off.

The comparison between eqs. (\ref{WLpert}) and (\ref{WLMM}) shows that, apart from the overall framing--one factor, at third order in the couplings a mismatch appears due to a non--trivial imaginary contribution in the matrix model result (framing -1) that is absent in the result obtained in ordinary perturbation theory (framing zero). Requiring consistency between the two results leads to the conclusion that the non--trivial term in (\ref{WLMM}) has to be ascribed to a framing effect. In the next section we are going to prove that this is indeed the case, being this contribution associated to a higher--order correction to the framing function coming from a non--vanishing finite two--loop correction to the gauge propagator.  As a by--product we also find strong evidence that statement (2) above is no longer true, since matching with the matrix model result works only if vertex--type diagrams contribute to framing in canceling unwanted terms from the propagator corrections.

\section{Gauge effective propagator in ABJ(M)}
 
As discussed above, we expect higher--order corrections to the framing function to come from non--trivial quantum corrections to the vector propagator. In this section we then concentrate on loop corrections to the $A_\mu$  propagator. We will perform the explicit calculation up to two loops and discuss the general structure at higher orders. A similar calculation can be easily applied also to $\hat{A}_\m$. 

Gauge and local Lorentz invariance require the quantum gauge propagator in momentum space to be of the form\footnote{Below
$ \Pi_e(p,\l_i)$  and $\Pi_o (p,\l_i)$ are exchanged with respect to the usual convention \cite{Dunne:1998qy} where the subscripts $e$ (even) and $o$ (odd) typically denote
the behavior under parity. We prefer, instead, to use these subscripts to indicate the loop order: $\Pi_e$ has an expansion in even powers of the coupling and $\Pi_o$ in odd ones, according to the normalization \eqref{genprop}.}
\bea
\label{genprop}
&& \langle A_\mu(p) A_\nu(-p) \rangle \equiv \frac{2\pi}{k} \Pi_{\m\n} (p,\l_i)  
\\
&& \Pi_{\m\n} (p,\l_i) =  \Pi_e(p,\l_i)\e_{\mu\nu \rho} p^{\rho} + \Pi_o(p,\l_i) \left( \d_{\mu \n} - \frac{p_\m p_\n}{p^2} \right) \non 
\eea
where $\Pi_e(p,\l_i), \Pi_o(p,\l_i)$ are power series in the two couplings $\l_1, \l_2$. 

In general, at a given order, the first tensorial structure will come out from diagrams proportional to an odd number of $\varepsilon$ tensors, whereas the second structure will be produced by diagrams proportional to an even number of $\varepsilon$. Recalling that  in Landau gauge $\varepsilon$ tensors come from vector propagators, gauge cubic vertices (see Appendix A) and traces of gamma matrices (from identity (\ref{id1}) and its generalizations) it is easy to prove that at even loop order the number of $\varepsilon$ tensors is always odd and conversely.
Therefore, the perturbative corrections to the gauge propagator in Landau gauge are proportional to the tensor structure $\e_{\mu\nu \rho} p^{\rho}$ at even loops and to the structure $( \d_{\mu \n} - \frac{p_\m p_\n}{p^2} )$ at odd loops. This is confirmed by the explicit expressions of the tree and one--loop propagators in eqs. (\ref{treevector}, \ref{1vector}). It follows that $\Pi_e(p,\l_i)$
and $\Pi_o(p,\l_i)$ are even and odd expansions in the couplings, respectively. 

According to the general discussion of the previous section we expect that the function $\Pi_e(p,\l_i)$, if not vanishing, will contribute to correct the framing function at higher orders. To give support to this expectation we then compute the first non--trivial contribution to $\Pi_e(p,\l_i)$, that is the two--loop gauge propagator.
 
\subsection{Feynman diagram computation}

We compute the two--loop gauge propagator in Landau gauge and in the planar limit using DRED scheme \cite{Siegel, Siegel:1980qs} that respects supersymmetry and gauge invariance. 
From previous calculations \cite{Chen:1992ee} we know that at this order there are no corrections to the gauge self--energy coming from the pure CS sector of the theory. Therefore it is sufficient to consider matter contributions.  

\begin{figure}[h]
\begin{center}
 \includegraphics[width=0.80\textwidth]{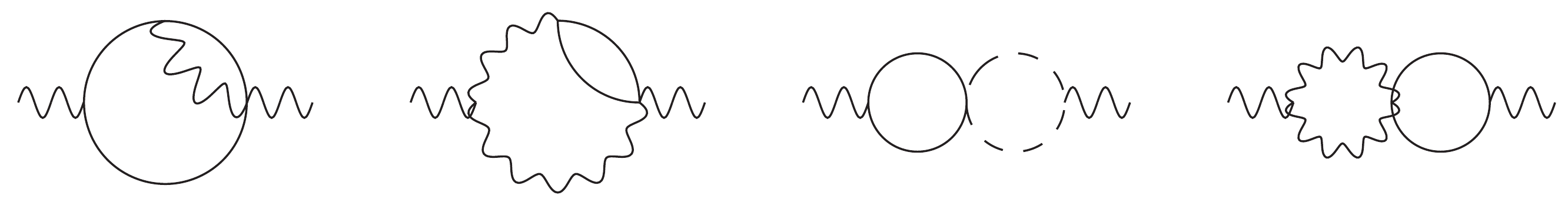}
\caption{Two-loop self energy diagrams  with an identically vanishing integral.}\label{vanishing}
\end{center}
\end{figure}

Taking into account that there is no one--loop correction to the scalar propagator and excluding diagrams that give rise to vanishing integrals (see Fig. \ref{vanishing})
  
\noindent  
the complete list of contributions to the vector self-energy at  two loops is given in Table \ref{tab},  where the color factors are also indicated.
\begin{table}
\begin{tabular}{ccccccc}
 & Diagram & Color & & & Diagram & Color \\[5mm]
(a) & \raisebox{-0.5cm}{\includegraphics[scale=0.5]{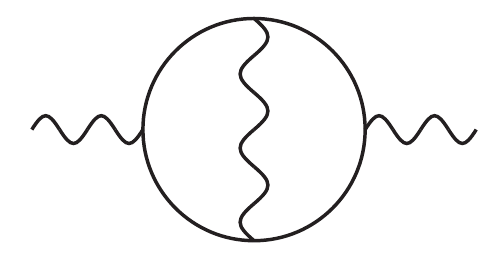}} & $N_2^2$& & \hspace{1.5cm} (e) & \raisebox{-0.5cm}{\includegraphics[scale=0.5]{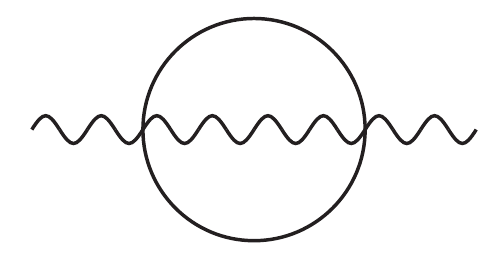}} & $2N_2^2-N_1N_2$ \\[5mm]
(b) & \raisebox{-0.5cm}{\includegraphics[scale=0.5]{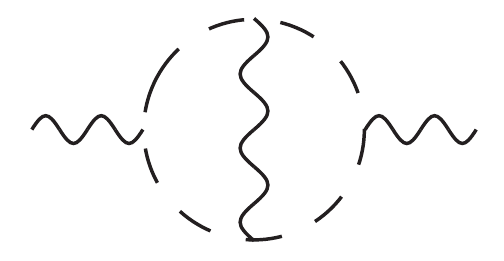}} & $N_2^2$ && \hspace{1.5cm} (f) & \raisebox{-0.5cm}{\includegraphics[scale=0.5]{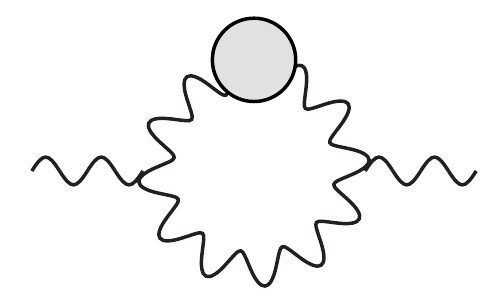}} & $N_1N_2$ \\[5mm]
(c) & \raisebox{-0.5cm}{\includegraphics[scale=0.5]{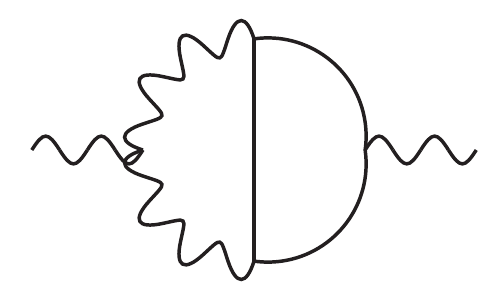}} & $N_1N_2$ &&  \hspace{1.5cm} (g) & \raisebox{-0.5cm}{\includegraphics[scale=0.5]{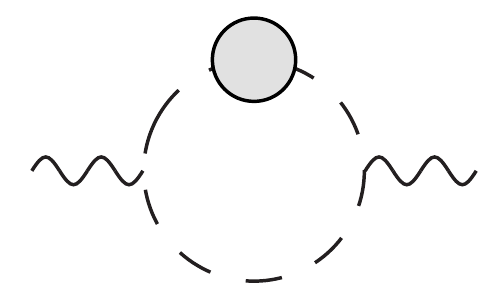}} & $N_1N_2-N_2^2$  \\[5mm]
(d) & \raisebox{-0.5cm}{\includegraphics[scale=0.5]{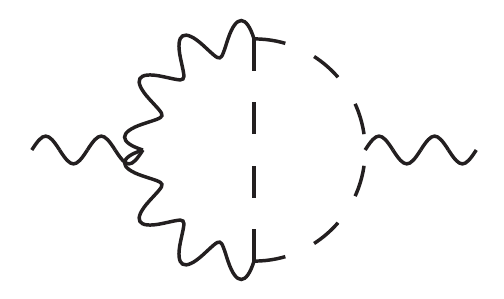}} & $N_1N_2$ && \hspace{1.5cm} (h) & \raisebox{-0.2cm}{\includegraphics[scale=0.5]{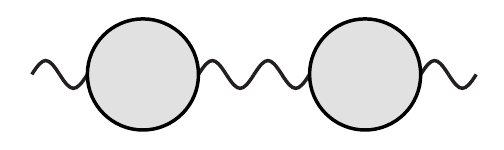}} & $N_2^2$ 
\end{tabular}
\caption{Two loop non vanishing contributions with correspondent associated color structure.}
\label{tab}
\end{table}
The corresponding integrals  are UV divergent except the last one. We regularize divergences by working in $d = 3 -2 \e$ dimensions. 

As a general strategy, we compute the second order contribution to $\Pi_e$ by contracting the two--loop propagator with the tensorial structure $[\varepsilon_{\mu\nu\rho} p^\rho]^{-1} = \tfrac{1}{2p^2} \varepsilon_{\mu\nu\rho} p^\rho$. Precisely, we write
\beq
\Pi_e^{(2)} (p) = \tfrac{1}{2p^2} \varepsilon_{\mu\nu\rho} p^\rho \, \Pi_{\m\n}^{(2)} (p) 
 \eeq
where $\Pi_{\m\n}^{(2)} (p) $ is the sum of the contributions in Table \ref{tab}. With this operation we trade the evaluation of the original tensor integrals with the evaluation of scalar integrals, so avoiding regularization ambiguities due to the contraction of three dimensional tensors with $d$ dimensional momenta coming from the regularized integrals.

We perform the calculation in momentum space and refer to Appendix A for the corresponding Feynman rules. For each diagram we just write the initial expression and the final result, adding few details only in the most complicated cases. All the results are given in terms of $G$ functions defined as
\beq
\int \frac{d^dk}{(2\pi)^{d}}\, \frac{1}{[k^2]^{a}[(k-p)^2]^{b} } \equiv \frac{1}{(p^2)^{a + b - d/2}} \, G_{a,b}
\eeq
with 
\beq
\label{G1}
G_{a,b} = \,\frac{1}{(4\pi)^{d/2}} \frac{\G(a + b - \tfrac{d}{2}) \G( \frac{d}{2} - a) \G( \frac{d}{2} - a) }{\G(a) \G(b) \G(d - a-b)}
\eeq

\vskip 20pt

\paragraph{\underline{Diagram (a):}}

The contribution of diagram (a) to the effective propagator (\ref{genprop}) is
\begin{align}\label{contra}
{\rm (a)} =  \left(\frac{2\pi}{k}\right)^3 \! 64N_2^{\,2} \, \, \frac{\varepsilon_{\mu\rho_1\rho_2}\varepsilon_{\rho_3\rho_4\rho_5} \varepsilon_{\rho_6\nu\rho_7}  p^{\rho_2} p^{\rho_4}p^{\rho_7}}{p^4} \int \frac{d^dl\, d^dk}{(2\pi)^{2d}}\, \frac{ k^{\rho_1}l^{\rho_3} k^{\rho_5} l^{\rho_6}}{k^2(k-p)^2(k-l)^2(l-p)^2l^2} 
\end{align}
We proceed as discussed above by contracting  with the tensorial structure  $ \varepsilon_{\mu\nu\rho} p^\rho$ and converting contracted pairs of $\varepsilon$ to combinations of delta functions. This allows to rewrite the string of momenta at numerator as a combination of scalar structures
\begin{equation}
p^2 \, (k \cdot l)  ( (l \cdot k) \, p^2- (l \cdot p) \, (k \cdot p))-k^2 p^2 (l^2 p^2-(l \cdot p)^2)+p^2\,l^2(k \cdot p)^2-p^2\, (k \cdot l) \, (k \cdot p) \,(l \cdot p)
\end{equation}
Completing the squares and discarding tadpoles, we can read the final contribution to the propagator  
\bea
{\rm (a)} &=&  \left(\frac{2\pi}{k}\right)^3\! N_2^{\,2}
\,\bigg(4\, G_{1,1}^2-\frac{16}{3}\,G_{1,1}G_{1,1/2+\epsilon}\bigg)\,
\frac{\varepsilon_{\mu\nu \rho}\,p^{\rho}}{(p^2)^{1+2\epsilon}}
\eea
with the $G$ functions defined in (\ref{G1}).

\paragraph{\underline{Diagram (b):}}

The contribution from diagram (b)  reads
\begin{align}
{\rm (b)} &= \left(\frac{2\pi}{k}\right)^3 \! 4N_2^{\,2} \, \, \frac{\varepsilon_{\mu\rho_1\rho_2}\varepsilon_{\rho_3\rho_4\rho_5}\varepsilon_{\rho_6\nu\rho_7}p^{\rho_2}p^{\rho_7}}{p^4} \int \frac{d^dl\, d^dk}{(2\pi)^{2d}}\, \frac{(k-l)^{\rho_5}k^{\rho_8}l^{\rho_{9}}(l-p)^{\rho_{10}}(k-p)^{\rho_{11}}}{k^2(k-p)^2(k-l)^2(l-p)^2l^2} \, \nonumber\\&\hspace{1cm} {\rm Tr}(\gamma^{\rho_1}\gamma^{\rho_8}\gamma^{\rho_3}\gamma^{\rho_9}\gamma^{\rho_6}\gamma^{\rho_{10}}
\gamma^{\rho_4}\gamma^{\rho_{11}}) 
\end{align}
This expression is again brought to scalar form by contraction with  $ \varepsilon_{\mu\nu\rho} p^\rho$. After working out some lengthy $\g-$algebra and completing the squares,  the final result can be expressed as 
\begin{align}
{\rm (b)} &= \left(\frac{2\pi}{k}\right)^3 \! N_2^{\,2} \, \left(4\, G_{1,1}^2 -\frac{16}{3}\, G_{1,1} G_{1,1/2+\epsilon}  \right)\, 
\frac{\varepsilon_{\mu\nu \rho}\,p^{\rho}}{(p^2)^{1+2\epsilon}}
\end{align}
and turns out to be identical to the contribution of diagram (a). 

\paragraph{\underline{Diagram (c):}} This diagram yields
\begin{equation}
{\rm (c)} = - \left(\frac{2\pi}{k}\right)^3 \! 64 N_1N_2 \, \, \frac{\varepsilon_{\mu\rho_1\rho_2}\varepsilon_{\rho_3\rho_4\rho_5} \varepsilon_{\rho_6\nu\rho_7}  p^{\rho_2} p^{\rho_4}p^{\rho_7}}{p^4} \int \frac{d^dl\, d^dk}{(2\pi)^{2d}}\, \frac{ k^{\rho_1}l^{\rho_3} k^{\rho_5} l^{\rho_6}}{k^2(k-p)^2(k-l)^2(l-p)^2l^2} 
\end{equation}
Besides an overall sign and the different color structure, this contribution is exactly the same as the one appearing in diagram (a). Exploiting the previous result we can write 
\begin{equation}
{\rm (c)}=   \left(\frac{2\pi}{k}\right)^3  \! N_1 N_2 
\,\bigg(\frac{16}{3}\,G_{1,1}G_{1,1/2+\epsilon}- 4\,G_{1,1}^2\bigg)\,
\frac{\varepsilon_{\mu\nu \rho}\,p^{\rho}}{(p^2)^{1+2\epsilon}}
\end{equation}

\paragraph{\underline{Diagram (d):}}  This diagram is the fermion counterpart of diagram (c) and gives
\begin{align}
{\rm (d)} &=  - \left(\frac{2\pi}{k}\right)^3 \! 8 N_1N_2 \,\, \varepsilon_{\mu\rho_1\rho_2} \varepsilon_{\rho_1\rho_3\rho_4}\,\varepsilon_{\rho_3\rho_5\rho_6}\, \varepsilon_{\rho_7\rho_4\rho_8}\, \varepsilon_{\rho_9\nu\rho_{10}}\, p^{\rho_2}\, p^{\rho_{10}} \nonumber\\& \int \frac{d^dl\, d^dk}{(2\pi)^{2d}}\, \frac{{\rm Tr}\left(\g^{\rho_5} \g^{\sigma_1} \g^{\rho_9} \g^{\sigma_2} \g^{\rho_7} \g^{\sigma_3}\right) k^{\sigma_1} (k-p)^{\sigma_2} (k-l)^{\sigma_3} (l-p)^{\rho_8} l^{\rho_6}}{k^2(k-p)^2(k-l)^2(l-p)^2l^2}
\end{align}
As usual  we multiply the external structure by $\varepsilon_{\mu\nu\rho}p^\rho$ and expand every pair of $\varepsilon$-tensors. After a lengthy computation, the numerator of the integral is reduced to the following scalar expression
\begin{equation}
2\, (k\cdot p)\,(k-p)\cdot (l-p)\,(k-l)\cdot l-
2\, p\cdot (k-p)\,(k\cdot l)\,(k-l)\cdot (l-p)
\end{equation}
Then, completing squares and discarding tadpoles we finally arrive to 
\begin{equation}
{\rm (d)}= \left(\frac{2\pi}{k}\right)^3  \! N_1 N_2 
\,\bigg(2\, G_{1,1}^2-\frac{8}{3}\,G_{1,1}G_{1,1/2+\epsilon}\bigg)\,
\frac{\varepsilon_{\mu\nu \rho}\,p^{\rho}}{(p^2)^{1+2\epsilon}}
\end{equation}

\paragraph{\underline{Diagram (e):}}

This diagram is quite easy to evaluate. After Wick contractions we obtain
\begin{align}
{\rm (e)} &=  - \left(\frac{2\pi}{k}\right)^3\! (16 N_2^{\,2} - 8 N_1N_2 ) \frac{\varepsilon_{\mu\rho_1\rho_2}\varepsilon_{\rho_1\rho_3\rho_4}\varepsilon_{\rho_3\nu\rho_5}p^{\rho_2}p^{\rho_5}}{p^4} \int \frac{d^dl\, d^dk}{(2\pi)^{2d}}\, \frac{l^{\rho_4}}{k^2(k+l-p)^2l^2} 
\end{align}
Following the previous steps, we immediately get a scalar integral from which we extract the contribution to the propagator
\begin{align}
{\rm (e)} & = \left(\frac{2\pi}{k}\right)^3 \left(N_2^{\,2} - \frac{1}{2} N_1N_2 \right)
\,\bigg(\frac{16}{3}\,G_{1,1}G_{1,1/2+\epsilon} \bigg)\,
\frac{\varepsilon_{\mu\nu \rho}\,p^{\rho}}{(p^2)^{1+2\epsilon}}
 \end{align}
 
\paragraph{\underline{Diagram (f):}} This diagram contains the one--loop correction to the gauge propagator (\ref{1vector})
which, once inserted in two possible ways in the vector bubble, leads to  
\begin{align}
{\rm (f)} & =  \left(\frac{2\pi}{k}\right)^3\! 8 N_1N_2 \, G_{1,1} \, \varepsilon_{\mu\rho_1\rho_2} \varepsilon_{\rho_1\rho_3\rho_4} \varepsilon_{\rho_5\rho_6\rho_7}\, \varepsilon_{\rho_7\rho_4\rho_9} \varepsilon_{\rho_6\nu\rho_8} p^{\rho_2}p^{\rho_8}\,\non \\ 
&\hspace{2cm}\times \int \frac{d^d k}{(2\pi)^d}\,  \frac{(k-p)^{\rho_9}}{(k-p)^2(k^2)^{1/2+\epsilon}} \left( \eta_{\rho_3\rho_5} - \frac{k_{\rho_3}k_{\rho_5}}{k^2}\right)
\end{align}
Proceeding as in the previous cases we obtain
\begin{align}
{\rm (f)} &= \,  \left(\frac{2\pi}{k}\right)^3\, N_1 N_2\,\bigg(\frac{10}{3}\,G_{1,1}G_{1,1/2+\epsilon} - 2\,G_{1,1}G_{1,3/2+\e} \bigg) \,
\frac{\varepsilon_{\mu\nu \rho}\,p^{\rho}}{(p^2)^{1+2\epsilon}}
\end{align}

\paragraph{\underline{Diagram (g):}} This diagram involves the 1-loop corrected fermion propagator given in (\ref{1fermion}). After taking into account the two inequivalent insertions in the fermion bubble we get
\begin{align}
{\rm (g)} &= - 8i \left(\frac{2\pi}{k}\right)^3\! (N_2^{\,2} - N_1N_2 ) \,  G_{1,1}\, {\rm Tr}\left(\gamma^{\rho_1}\gamma^{\rho_3}\gamma^{\rho_4}\right)\, \frac{ \varepsilon_{\mu\rho_1\rho_2} \varepsilon_{\rho_3\nu\rho_5}  p^{\rho_2} p^{\rho_5}}{p^4} 
\non \\
& \qquad \times \int \frac{d^dl}{(2\pi)^{d}}\, \frac{l^{\rho_4}}{l^2[(l-p)^2]^{1/2+\epsilon}} 
\end{align}
This can be  easily $\varepsilon$-contracted and manipulated to get the final contribution to the propagator
\begin{align}
{\rm (g)}& = \left(\frac{2\pi}{k}\right)^3(N_2^{\,2} -  N_1N_2 )
\,\bigg(\frac{16}{3}\,G_{1,1}G_{1,1/2+\epsilon} \bigg)\,
\frac{\varepsilon_{\mu\nu \rho}\,p^{\rho}}{(p^2)^{1+2\epsilon}}
 \end{align}
\paragraph{\underline{Diagram (h):}} The 1P--reducible contribution can be immediately derived from the 1-loop correction to the vector self energy, eq. (\ref{1vector}), and gives 
\begin{equation}\label{non1PI}
{\rm (h)} =  -\left(\frac{2\pi}{k}\right)^3 \! N_2^{\,2}
\,16\, G_{1,1}^2 \,
\frac{\varepsilon_{\mu\nu \rho}\,p^{\rho}}{(p^2)^{1+2\epsilon}}
\end{equation}

\subsection{The two-loop result}

We are now ready to sum the previous contributions and obtain the two--loop correction to the gauge propagator due to matter interactions. In terms of $G$ functions we have
\bea 
&& \frac{2\pi}{k} \Pi^{(2)}_{\m\n} (p,\l_i) = 
\\
&~& \quad - \left(\frac{2\pi}{k}\right)^3 \, \left[ 8  N_2^2 \, G_{1,1}^2  
+  2  N_1 N_2 \, \left( G_{1,1}^2 +G_{1,1} G_{1,1/2+\epsilon} + G_{1,1} G_{1,3/2+\epsilon} \right) \right] \, \frac{\varepsilon_{\mu\nu \rho}\,p^{\rho}}{(p^2)^{1+2\epsilon}}
\non 
\eea
Using the explicit expansions 
\begin{align}
G_{1,1}^2  & =
\frac{1}{64} + {\cal O}(\e)  \\
G_{1,1}G_{1,1/2+\epsilon} & =   \,\frac{e^{-2\gamma_E\epsilon}}{(4\pi)^{3-2\e}} \pi
\,\left(\frac{1}{\e} + 6 + {\cal O}(\e) \right) \\
G_{1,1}G_{1,3/2+\epsilon} & =   \,\frac{e^{-2\gamma_E\epsilon}}{(4\pi)^{3-2\e}} \pi
\,\left(-\frac{1}{\e} + 2 + {\cal O}(\e) \right)
\end{align}
one can immediately realize that the $1/\e$ divergences cancel and for $\e \to 0$ we are left with a finite result given by
\beq
\label{2Lprop}
 \frac{2\pi}{k}\Pi^{(2)}_{\m\n} (p,\l_i) \equiv   \frac{2\pi}{k} \Pi^{(2)}_e(p,\l_i) \e_{\mu\nu \rho} \, p^{\rho} = -\frac{\pi^3 }{k^2} \left[ N_2^2 +   N_1 N_2 \left( \frac{1}{4}  + \frac{2}{\pi^2} \right) \right]  \frac{ \varepsilon_{\mu\nu \rho}\,p^{\rho} }{p^2}
\eeq
Since in dimensional reduction two--loop contributions from the gauge and ghost sectors cancel each other \cite{Chen:1992ee}, expression (\ref{2Lprop}) is the complete two--loop correction to the gauge propagator in ABJ(M) theory. 
 
Transforming back to configuration space we obtain
\beq
\label{2loop}
\langle A_\mu(x) A_\nu(y) \rangle^{(2)} = -\left( \frac{2 \pi i }{k} \right) \left[ \l_2^2 +   \l_1 \l_2 \left( \frac{1}{4}  + \frac{2}{\pi^2} \right) \right] \frac{\G(\frac32 - \e)}{4 \pi^{-\frac12 -\e}} \, \varepsilon_{\mu\nu \rho}
\, \frac{(x-y)^\rho}{[(x-y)^2]^{\frac32 -\e}}
\eeq
We stress the two--loop finiteness of the gauge propagator. This is a remarkable property of ABJ(M) theories, as in general for less supersymmetric CS-matter theories a non--trivial renormalization of the gauge connection enters already at two loops \cite{Chen:1992ee} together with a renormalization of the CS coupling in such a way that the beta function vanishes \cite{Kapustin:1994mt}.   
At this order this peculiar property of ABJ(M) is due to the presence of two gauge groups with opposite CS levels and the particular configuration of matter in bifundamental representation, but not to the details of the self--interactions in the matter sector. We expect them to play a role at higher loops where it would be nice to check whether the finiteness of the gauge propagator is still valid. 

Since these theories are expected to have vanishing beta functions \cite{Kapustin:1994mt}, the lack of renormalization of the gauge self--energy at two loops necessarily implies that no renormalization of the CS cubic interaction (and then of the CS coupling) will arise at this order.

\section{Framing at three loops for the $1/6$ BPS WL}

As discussed in section 2, the $\varepsilon_{\mu\nu\rho} p^\rho$ part of the gauge propagator (\ref{genprop}) is responsible for the emergence of the framing factor. The tree--level 
propagator contributes to the framing function only with terms proportional to powers $\l_1^n$ that are known to exponentiate the $1/k$ contribution (pure CS sector). In addition,  higher--order corrections to the framing function may come from non--trivial corrections to the $\Pi_e$ function when the effective propagator is used in
\beq
\label{WL2loop}
\langle \mathcal{W} \rangle \rightarrow \frac{1}{N_1} \, (-i)^2  \int_0^{2\pi} d\tau_1 \int_0^{\tau_1} d\tau_2 \, \dot{x}_1^\mu \, \dot{x}_2^{\nu} \, \langle A_\mu(x_1) A_\nu(x_2) \rangle  
\eeq
where the integral is framed according to the prescription (\ref{frame2}). 

We discuss this effect in details. Since at two loops the corrections to the propagator are proportional to two different color structures, we will consider the two cases separately.

\subsection{Color structure $\l_1 \l_2^2$}

We begin by inserting in eq. (\ref{WL2loop}) the $\l_2^2$ term of the two--loop propagator (\ref{2loop}).  Using point--splitting regularization as done at tree level, when the regularization parameter is removed we are left with the following third order frame--dependent contribution
\begin{equation}
\langle \mathcal{W} \rangle^{(3)}|_{\l_1\l_2^2} = i\frac{\pi^3}{2} \l_1 \l_2^2 \, \chi(\G,\G_f) 
\end{equation}
where $\chi(\G,\G_f)$ is the linking number of the two closed paths defined in eq. (\ref{gauss}). In particular, setting $\chi(\G,\G_f) =-1$, the result perfectly matches the third order contribution in the localization result \cite{Drukker:2010nc} (see eq. (\ref{WLMM})) and elucidates its framing origin. 

We remark that the choice of DRED scheme has been crucial to get rid of lower transcendentality constants which are not present in the localization result.

The factorization theorem for reducible diagrams proved in the pure CS case \cite{Labastida} should still work in the presence of matter and with non--trivial quantum corrections to the gauge propagator. It then follows that multiple insertions of collapsible gauge propagators corrected at two loops (ladder--type topologies as in Fig. \ref{expo} with subdiagrams now containing also matter) lead to the exponentiation of the framing function, now corrected at order $\l_1 \l_2^2$.

\begin{figure}[h]
\begin{center}
 \includegraphics[width=0.60\textwidth]{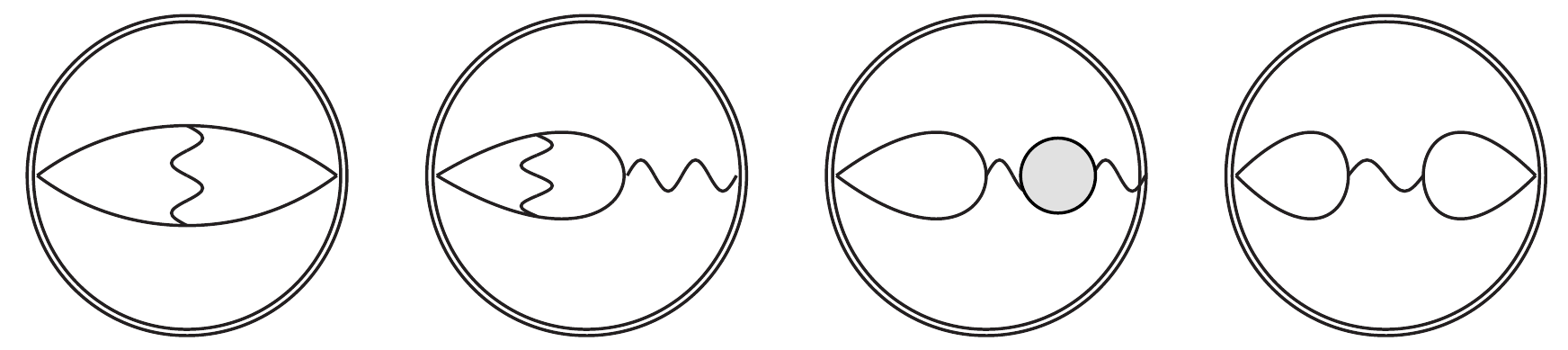}
\caption{Two-loop diagrams proportional to the color structure $\l_1 \l_2^{2}$ not containing collapsible gauge propagators.  }\label{color}
\end{center}
\end{figure}

At this order there can be other potential sources of framing proportional to this color factor not ascribable to propagator--type diagrams.  There are indeed further diagrams, the ones in Fig. \ref{color}, that together with propagator diagrams already discussed, give the {\it complete} set of contributions proportional to $\l_1 \l_2^2$. However, a close inspection reveals that all these diagrams vanish identically either because they are proportional to vanishing integrals or because they give rise to the trace of odd powers of the matrix $M$.

\noindent
In conclusion, as long as the $\l_1 \l_2^2$ correction to framing is concerned, at framing -1 we can write 
\beq
\label{partial}
\langle \mathcal{W} \rangle = e^{i \pi \left( \l_1 - \frac{\pi^2}{2} \l_1\l_2^2 + {\cal O}(\l^5) \right)} \left( 1 - \frac{\pi^2}{6} (\l_1^2 - 6 \l_1\l_2)  + {\cal O}(\l^4) \right) 
\eeq
where the framing factor $e^{- i\frac{\pi^3}{2} \l_1\l_2^2}$ is the result of resumming an infinite subclass of diagrams, that is the ones containing multiple collapsible gauge propagators corrected at two loops.

\subsection{Color structure $\l_1^2 \l_2$}

\begin{figure}[h]
\begin{center}
 \includegraphics[width=0.35\textwidth]{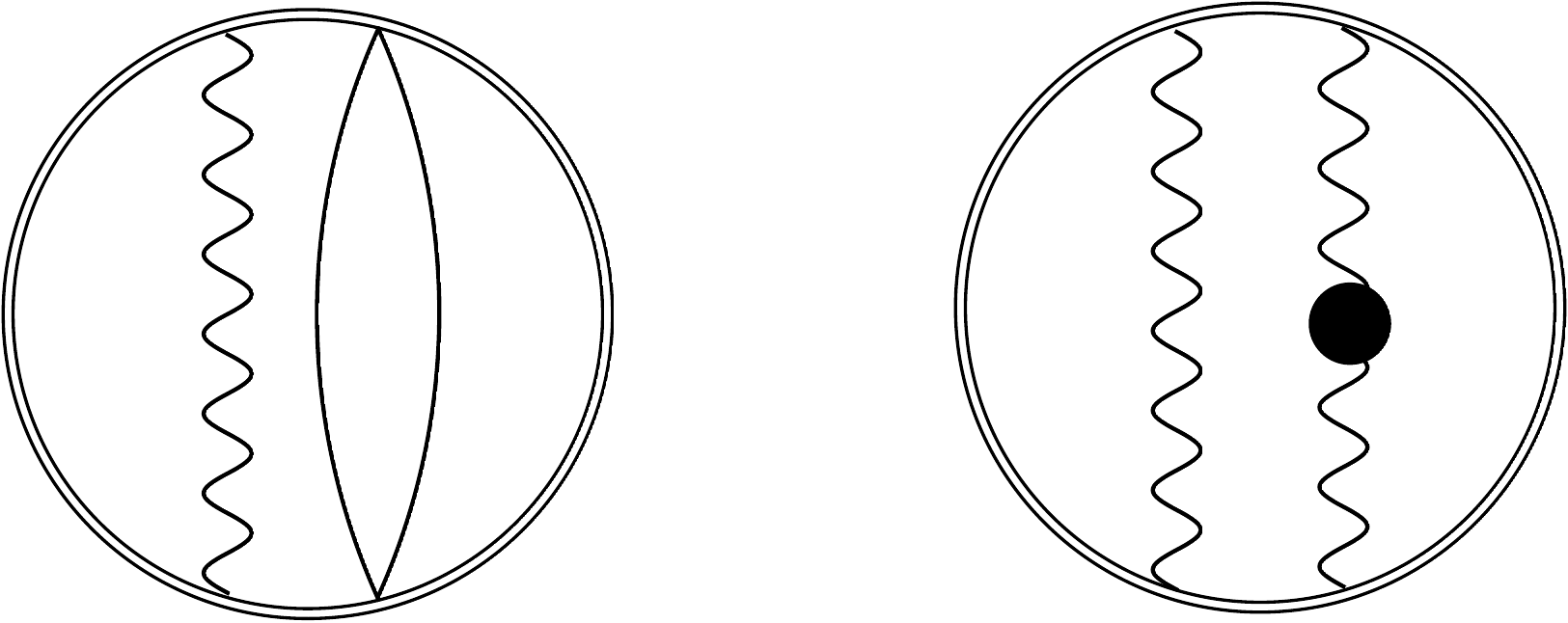}
\caption{Framing dependent diagrams with $\l_1^2 \l_2$ color structure which reconstruct the standard phase factor.}\label{fact}
\end{center}
\end{figure}

The result from localization, eq. (\ref{WLMM}), does not exhibit any correction proportional to the color structure $\l_1^2 \l_2$ once the standard framing phase $e^{\pi i \l_1}$ is  factorized. Expanding the phase in (\ref{WLMM}) we obtain a term
\begin{equation}\label{fac}
\langle \mathcal{W} \rangle^{(3)}|_{\l_1^2\l_2} = - i \pi^3 \l_1^2 \l_2  
\end{equation}
which should  be reproduced diagrammatically. If factorization of the framing works as in the pure CS case, we expect that this contribution should be entirely due to the diagrams in Fig. \ref{fact} (and their permutations) containing one collapsible tree--level gauge propagator. Indeed, these diagrams represent the factorized interference between the two--loop matter diagrams and the gauge vector exchange.  
We have explicitly checked that these diagrams reproduce exactly contribution (\ref{fac}). This is a non--trivial check that the exponentiation of the pure CS framing works also in the presence of matter.

\begin{figure}[h]
\begin{center}
 \includegraphics[width=0.55\textwidth]{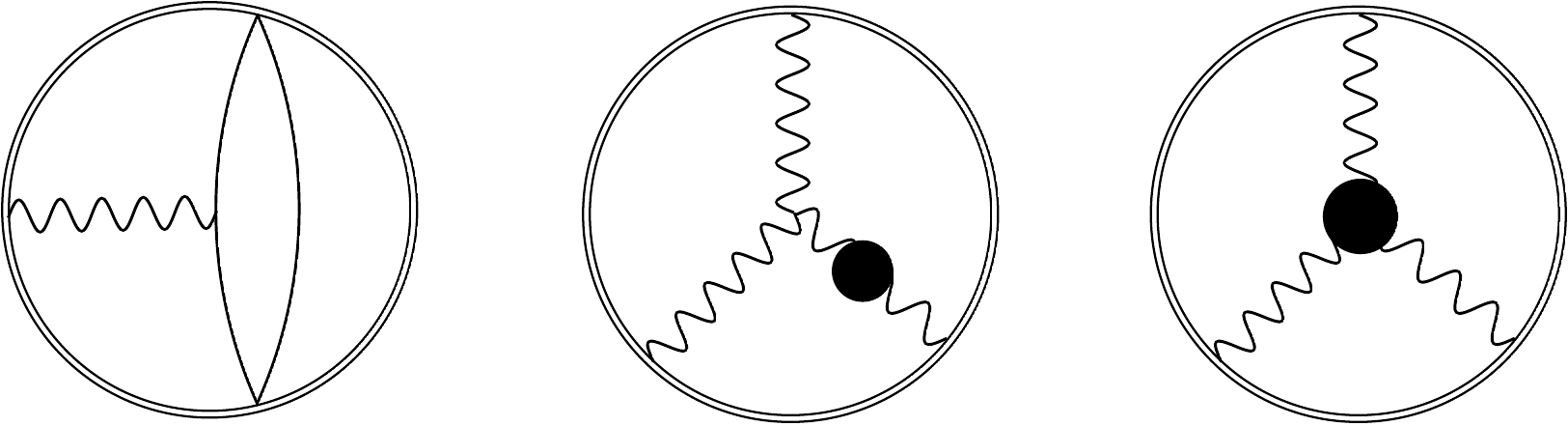}
\caption{List of possible framing dependent diagrams with color  $\l_1^2 \l_2$ and no  collapsible vector propagators.}\label{vertex}
\end{center}
\end{figure}

Once this is taken into account the matrix model predicts no further contributions to the $\l_1^2 \l_2$ color sector.
However,  the two--loop propagator (\ref{2loop}) contains a correction proportional to $\l_1 \l_2$ and its insertion in eq. (\ref{WL2loop}) gives rise to a non--vanishing third--order correction to $\langle \mathcal{W} \rangle$ given by
\begin{equation}\label{plus}
  i\frac{\pi^3}{2} \l_1^2 \l_2 \, \left( \frac{1}{4}  + \frac{2}{\pi^2} \right) \,  \chi(\G,\G_f) 
\end{equation}
This expression not only contains weird lower transcendentality terms, but would be also in contrast with the matrix model result. The combination of these two unexpected outputs leads to the expectation that in this sector framing dependent contributions come also from diagrams with no collapsible propagators, precisely the ones listed in Fig. \ref{vertex}. Matching with the matrix model prediction necessarily implies that these vertex-type diagrams should cancel exactly contribution (\ref{plus}) coming from the propagator. 

A full analytical computation of these diagrams is complicated.
This entails both solving the internal interaction integrals and then analyzing the behavior of the integrand under integration on framing contours.
A complete analysis is beyond the scopes of this paper.
Here, we restrict to a sanity check of our conjecture, testing whether some of these diagrams are in fact able to develop a framing dependence.

\begin{figure}
\begin{center}
\includegraphics[width=4.cm]{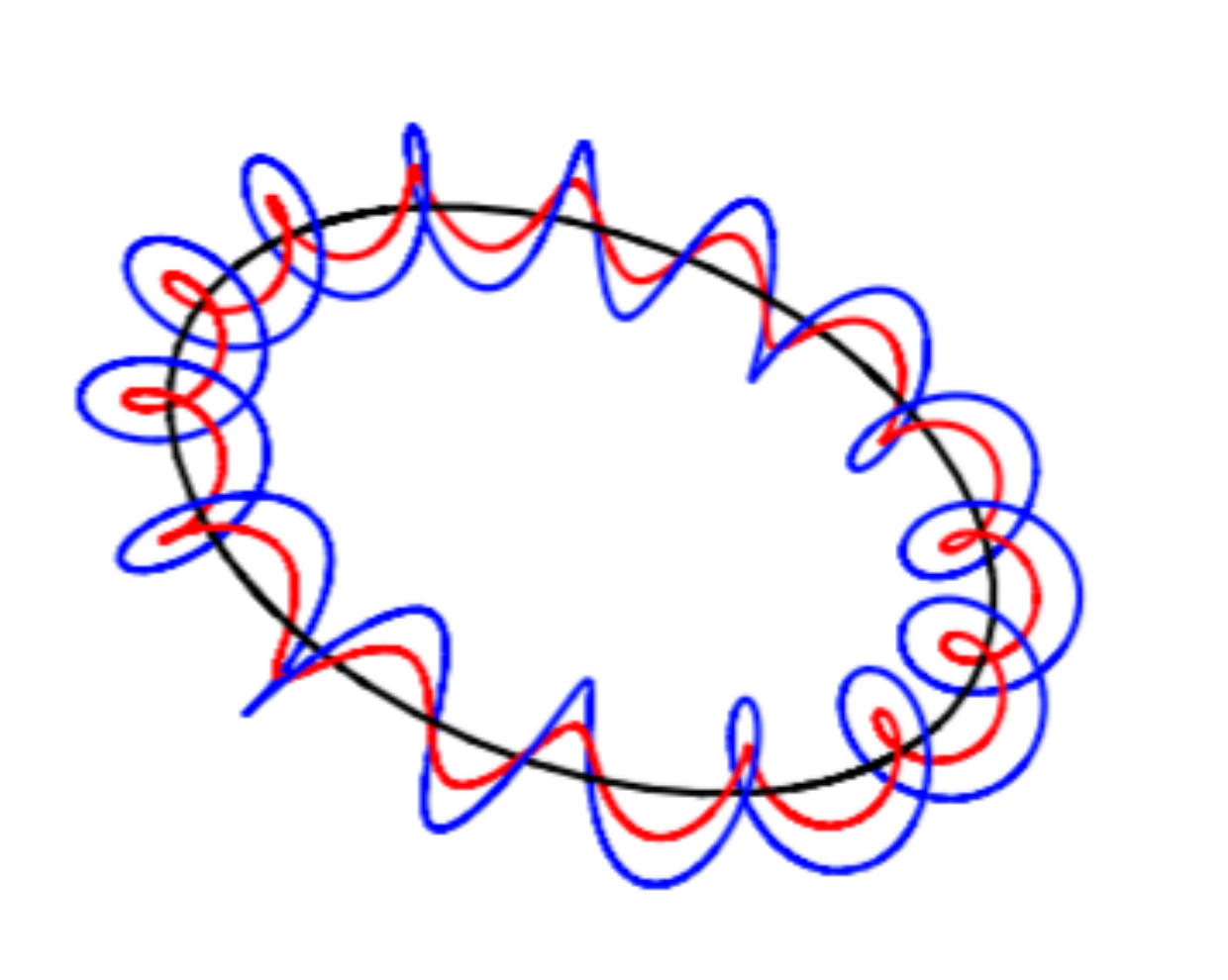}
\caption{A cartoon of framing contours.}
\label{framingfig}
\end{center}
\end{figure}

The simplest diagram to compute is the first one in Fig. \ref{vertex}.
After performing Feynman rules algebra we obtain a result proportional to (here we use $x_i \equiv x(\tau_i)$)
\begin{equation}
\label{vertexint}
\raisebox{-0.8cm}{\includegraphics[width=2.cm]{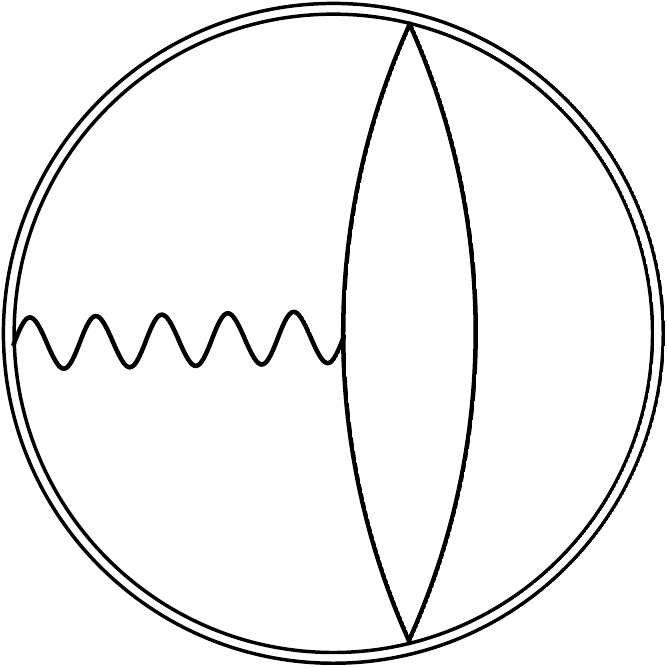}}
 \propto i\hspace{-0.5cm} \int\limits_{2\pi>\tau_1>\tau_2>\tau_3>0}\hspace{-1.0cm} d\tau_1 d\tau_2 d\tau_3 \, \varepsilon_{\mu\nu\rho}\, x_{12}^{\nu} x_{23}^{\rho}
\frac{\dot x_2^{\mu}\, |\dot x_1||\dot x_3|\left(|x_{12}|+|x_{13}|-|x_{23}|\right)}{|x_{12}||x_{13}|^2|x_{23}|(|x_{12}||x_{13}|+x_{12}\cdot x_{13})} + cyclic
\end{equation}
The latter multiple integrals possess a singularity for coincident points which can potentially cause a framing dependence.
This is however nontrivial to establish analytically.
Alternatively, we provide numerical evidence that this integral can be framing dependent.
We evaluate the contour integral along a simple framing contour consisting of a toroidal helix of infinitesimal radius $\a$ winding around the original circular path
\begin{equation}\label{eq:contour}
x_i(\tau_i) = \{ 0, \cos\tau_i, \sin\tau_i  \} + \a \, (i-1) \,  \{ \sin n \tau_i, \cos n\tau_i \cos\tau_i, \cos n\tau_i\sin\tau_i \}
\end{equation}
 A magnified example is shown in Fig. \ref{framingfig}. The equation above indicates that the prescription for multiple point-splitting consists of shifting the contours with the same vector field, but different integer multiples of the magnitude, in such a way that all paths have the same linking number pairwise \cite{GMM}.

\begin{figure}
\begin{center}
\includegraphics[scale=0.6]{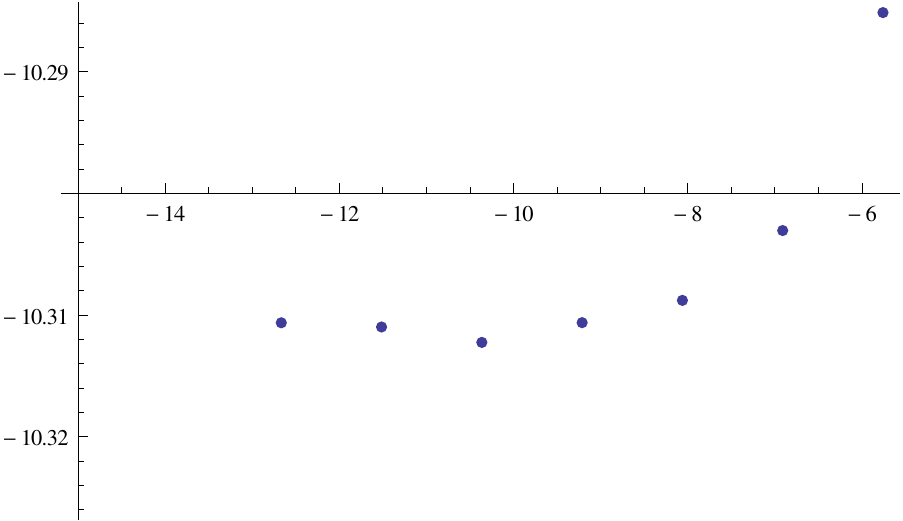}\hspace{1cm}
\includegraphics[scale=0.6]{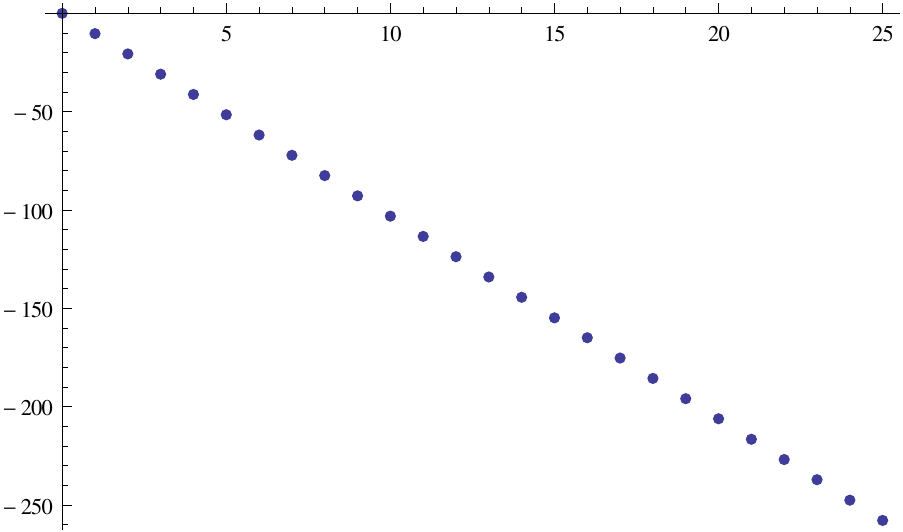}
\caption{Left: Dependence of integral (\ref{vertexint}) on $\log \a$ at fixed $n=1$: It approaches a finite value for $\a \to 0$. Right: Linear dependence on the framing number at fixed $\a=10^{-5}$.}
\label{scalar-gauge_plot}
\end{center}
\end{figure}

As a first check we study the behavior of the integral for fixed $n$ in the limit of vanishing $\a$. For $n=0$, namely trivial framing, the integrand vanishes identically. 
If the integral were to be framing independent, we would expect its value to tend to 0 even for generic $n$. 
On the contrary our numerical evaluation suggests that this is not the case, as the limit of vanishing $\a$ is finite but not zero when $n \neq 0$. In Fig. \ref{scalar-gauge_plot}  (left) we show an example of this limit for $n=1$.

As a second check we examine the dependence of the integral on the framing number $n$ for fixed and sufficiently small $\a$. 
 
We find that the dependence is linear, as the plot of Fig. \ref{scalar-gauge_plot} (right) indicates.
This is somehow in agreement with the expectation that this diagram could eventually contribute to the cancellation of \eqref{plus} where for the contour coordinates (\ref{eq:contour}) we have $\chi(\G, \G_f) =-n$. 

A similar numerical analysis can be performed on some pieces of the second diagram of Figure \ref{vertex} where the internal integral can be solved exactly. It exhibits the same finiteness properties in the $\a \rightarrow 0$ limit and a linear dependence on the framing number, as described above.

We stress that this analysis is incomplete and, moreover, it misses one crucial aspect. In fact, it does not address the question of whether the integrals we evaluate are metric dependent or not, that is if they depend only on the framing number or also on the particular shape of the framing contour. It is conceivable that the integrals of the various diagrams are individually metric and framing dependent, but that the sum only depends on the linking number of the framing contour.
The mechanism for this to occur is not clear and deserves further investigation.

\section*{Acknowledgements}

We acknowledge interesting discussions with Nadav Drukker, Marcos Marino and Diego Trancanelli. 
The work has been supported in part by INFN and MPNS--COST Action MP1210 ``The String Theory Universe".
The work of MB was supported in part by the Science and Technology Facilities Council Consolidated Grant ST/L000415/1 \emph{String theory, gauge theory \& duality}.

\vfill
\newpage

\appendix

\section{Conventions and Feynman rules}\label{sec:conventions}

In euclidean space the ${\cal N} = 6$ supersymmetric Chern--Simons--matter theory with gauge group $U(N_1)_k \times U(N_2)_{-k}$ \cite{ABJM, ABJ}  is described by the action  
\beq
\label{action}
S = S_{CS} + S_{gf} + S_{matter} 
\eeq
\bea
\label{CS}
S_{CS} &=& -i \frac{k}{4\pi}\int d^3x\,\varepsilon^{\mu\nu\rho} \Big[ \Tr \left( A_\mu\partial_\nu A_\rho+\frac{2}{3} i A_\mu A_\nu A_\rho \right)
 \\
&~& \qquad \qquad \qquad \qquad \quad - \Tr \left(\hat{A}_\mu\partial_\nu 
\hat{A}_\rho+\frac{2}{3} i \hat{A}_\mu \hat{A}_\nu \hat{A}_\rho \right) 
\Big]
\non \\
S_{gf} &=& \frac{k}{4\pi} \int d^3x \, \Tr \Big[ \frac{1}{\xi}  (\pa_\mu A^\mu)^2 + \pa_\mu \bar{c} D^\mu c  - 
\frac{1}{\xi} ( \pa_\mu \hat{A}^\mu )^2 - \pa_\mu \bar{\hat{c}} D^\mu \hat{c} \Big] 
\\
S_{matter} &=& \int d^3x \, \Tr \Big[ D_\mu C_I D^\mu \bar{C}^I + i \bar{\psi}^I \g^\mu D_\mu \psi_I \Big]  + S_{int}
\label{kin} 
\eea
where $S_{int}$ includes Yukawa vertices and sextic scalar interactions which are not needed at our perturbative order. Here $(C_I)^j_{\; \hat{j}}$ ($(\bar{C}^I)^{\hat{j}}_{\; j}$), $I=1, \cdots  4$,  are  four matter scalars in the  bifundamental (antibifundamental) representation of the gauge group, 
and $(\bar{\psi}^I)^j_{\; \hat{j}}$ ($(\psi_I)^{\hat{j}}_{\; j}$) are the corresponding fermions. Vector fields $A_\mu \equiv A_\mu^a T^a$ and $\hat{A}_\mu \equiv \hat{A}_\mu^a \hat{T}^a$ are the gauge potentials of the $U(N_1)$ and $U(N_2)$ groups respectively, with $\Tr (T^a T^b ) = \d^{ab}, \Tr (\hat{T}^{\hat{a}} \hat{T}^{\hat{b}} ) = \d^{\hat{a} \hat{b}}$.

\vskip 10pt
\noindent
Covariant derivatives are  defined as
\bea
\label{covariant}
D_\mu C_I &=& \pa_\mu C_I + i A_\mu C_I - i C_I \hat{A}_\mu \quad \qquad  ~ D_\mu \bar{\psi}^I  = \pa_\mu \bar{\psi}^I + i A_\mu \bar{\psi}^I - i \bar{\psi}^I \hat{A}_\mu
\non \\
D_\mu \bar{C}^I &=& \pa_\mu \bar{C}^I - i \bar{C}^I A_\mu + i \hat{A}_\mu \bar{C}^I  \quad \qquad D_\mu \psi_I = \pa_\mu \psi_I - i \psi_I A_\mu + i \hat{A}_\mu \psi_I 
\eea
Euclidean Clifford algebra  $\{ \g^\mu , \g^\nu \} = 2 \d^{\mu\nu}$ is explicitly realized by 
\beq
(\g^\mu)_\a^{\; \, \b} = \{ -\s^3, \s^1, \s^2 \}
\eeq
Spinorial indices are lowered and raised as $(\g^\mu)^\a_{\; \, \b} = \varepsilon^{\a \g}  (\g^\mu)_\g^{\; \, \d} \varepsilon_{\b \d}$, where $\varepsilon^{12} = \varepsilon_{21} = 1$. We conventionally choose to write the spinorial indices of chiral fermions always up, while the ones of antichirals always down
For instance, in (\ref{kin}) we read $ \bar{\psi}^I \g^\mu D_\mu \psi_I \equiv \bar{\psi}^I_\a (\g^\mu)^\a_{\; \, \b}  D_\mu \psi_I^\b$.

\vskip 10pt
\noindent
Products of gamma matrices can be easily sort out using the basic identity  
\beq 
\label{id1}
\g^\mu \g^\nu = \d^{\mu \nu} \mathbb{I} - i \varepsilon^{\mu\nu\rho} \g^\rho
\eeq

\noindent
From the action (\ref{action}), working in dimensional regularization ($d=3-2\e$) and in Landau gauge we obtain the following Feynman rules in configuration and momentum space

\vskip 15pt
\noindent
\underline{Vector propagators}  
\bea
\label{treevector}
&& \langle A_\mu^a (x) A_\nu^b(y) \rangle^{(0)} =  \d^{ab}   \, \left( \frac{2\pi i}{k} \right) \frac{\G(\frac32-\e)}{2\pi^{\frac32 -\e}} \varepsilon_{\mu\nu\rho} \frac{(x-y)^\rho}{[(x-y)^2]^{\frac32 -\e} }
\\
 &~&   \qquad \qquad \qquad \quad =\d^{ab}   \int \frac{d^dp}{(2\pi)^d} \frac{2\pi}{k} \,\varepsilon_{\mu\nu\rho} \,\frac{p^\rho}{p^2} \, e^{ip(x-y)}
\non \\
&& \langle A_\mu^a (x) A_\nu^b(y) \rangle^{(1)} = \d^{ab}   \left( \frac{2\pi }{k} \right)^2 N_2 \frac{\G^2(\frac12-\e)}{4\pi^{3 -2\e}} 
\left[ \frac{\d_{\mu\nu}}{ [(x- y)^2]^{1-2\e}} - \pa_\mu \pa_\nu \frac{[(x-y)^2]^{2\e}}{4\e(1+2\e)} \right]  \non  \\
&~ &   \qquad \qquad \qquad \quad = \d^{ab}     \left( \frac{2\pi }{k} \right)^2 N_2 \,   \frac{\G^2(\frac12-\e) \G(\frac12 + \e)}{\G(1-2\e) 2^{1-2\e} \pi^{\frac32 -\e}} 
\, \int \frac{d^dp}{(2\pi)^d} \frac{e^{ip(x-y)}}{(p^2)^{\frac12 + \e}} \left( \d_{\mu\nu} - \frac{p_\mu p_\nu}{p^2} \right)
\label{1vector}
\non \\
\eea
At tree--level the $\hat{A}$ propagator is minus the $A$ one, whereas at one loop it is the same but with $N_2$ replaced by $N_1$. 

\vskip 15pt
\noindent
\underline{Scalar propagator}
\bea
\label{scalar}
\langle (C_I)_i^{\; \hat{j}} (x) (\bar{C}^J)_{\hat{k}}^l(\; y) \rangle^{(0)}  = \d_I^J \d_i^l \d_{\hat{k}}^{\hat{j}} \, \frac{\G(\frac12 -\e)}{4\pi^{\frac32-\e}} 
\, \frac{1}{[(x-y)^2]^{\frac12 -\e}}
= \d_{\hat J}^{\hat I} \d_i^l \d_{\hat{k}}^{\hat{j}} \, \int \frac{d^dp}{(2\pi)^d} \frac{e^{ip(x-y)}}{p^2}
\non \\
\eea
The one--loop correction is vanishing.

\vskip 15pt
\noindent
\underline{Fermion propagator}
\bea
\label{treefermion}
&& \langle (\psi_I^\a)_{\hat{i}}^{\; j}  (x) (\bar{\psi}^J_\b )_k^{\; \hat{l}}(y) \rangle^{(0)} = - i \, \d_I^J \d_{\hat{i}}^{\hat{l}} \d_{k}^{j} \, 
\frac{\G(\frac32 - \e)}{2\pi^{\frac32 -\e}} \,  \frac{(\g^\mu)^\a_{\; \, \b} \,  (x-y)_\mu}{[(x-y)^2]^{\frac32 - \e}}
\\
&~& \qquad \qquad \qquad \quad \quad \quad  ~ =   -  \, \d_I^J \d_{\hat{i}}^{\hat{l}} \d_{k}^{j} \,    \int \frac{d^dp}{(2\pi)^d} \, (\g^\mu)^\a_{\; \, \b}  \frac{p_\mu}{p^2} \, e^{ip(x-y)}
\non \\
&& 
\\
\label{1fermion}
&& \langle (\psi_I^\a)_{\hat{i}}^{\; j}  (x) (\bar{\psi}^J_\b)_k^{\; \hat{l}}(y) \rangle^{(1)} = - \left( \frac{2\pi i}{k} \right) \,  \d_I^J \d_{\hat{i}}^{\hat{l}} \d_{k}^{j} \,  \, \d^\a_{\; \, \b}
\, (N_1-N_2) \frac{\G^2(\frac12 - \e)}{16 \pi^{3-2\e}} \, \frac{1}{[(x-y)^2]^{1 - 2\e}}  \non \\
\\
&~&  = - \left( \frac{2\pi i}{k} \right) \,\d_I^J \d_{\hat{i}}^{\hat{l}} \d_{k}^{j} \,  \, \d^\a_{\; \, \b} \, (N_1-N_2) \, 
\frac{\G^2(\frac12 - \e) \G(\frac12 + \e)}{\G(1 - 2\e) (4\pi)^{\frac32 - \e}} \, \int \frac{d^dp}{(2\pi)^d} \frac{e^{ip(x-y)}}{(p^2)^{\frac12 +\e}}
\non
\eea

\section{1/6 BPS WL: Expansion of Matrix Model result}\label{expansion}
 
From the matrix model description \cite{Marino:2009jd, Drukker:2010nc} it is possible to read the perturbative expansion of the expectation value of the 1/6 BPS Wilson loop. Here we give the first few terms of the expansion factorizing  the standard phase as in pure CS models 
   \begin{align}
\langle \mathcal{W} \rangle = & e^{\pi i \lambda_1} \left( 1-\frac{\pi^2}{6} (\l_1^2 - 6 \l_1 \l_2)  - i \frac{\pi^3 }{2} 
\l_1 \l_2^2 +  \frac{\pi^4 }{120} (
\l_1^4-10\l_1^3 \l_2-20\l_1\l_2^3 )\right. \non \\ &  + i \frac{\pi^5 }{24} 
\l_1 \l_2^2\, (\l_1^2+\l_2^2) +   \frac{\pi^6}{5040} (
-\l_1^6 + 14  \l_1^5 \l_2+ 700  \l_1^3 \l_2^3+42  \l_1 \l_2^5) \non \\ & - i \frac{\pi^7}{720} 
\l_1 \l_2^2 \,(\l_1^4+70\l_1^2\l_2^2 +\l_2^4) \non \\ & + \frac{\pi^8}{362880} (
\l_1^8  -18 \l_1^7 \l_2-7728  \l_1^5 \l_2^3 -56700  \l_1^4 \l_2^4-22932\l_1^3 \l_2^5-72 \l_1 \l_2^7) + \dots \bigg) 
\end{align}
Following our analysis of the perturbative corrections of the framing factor,  it might be useful to rewrite this result factorizing a generalized phase which multiplies a real function of the couplings
\begin{align}
\langle \mathcal{W} \rangle = &  e^{i \big[ \pi \l_1 - 
\frac{\pi^3}{2} \l_1 \l_2^2 + \frac{\pi^5}{24} \big(-
       \l_1^3 \l_2^2 +12
     \l_1^2 \l_2^3 + 
     \l_1 \l_2^4\big) +  \frac{\pi^7}{720} \big(-3
        \l_1^5 \l_2^2 + 
   60 \l_1^4 \l_2^3 - 425
     \l_1^3 \l_2^4 - 
   90  \l_1^2 \l_2^5 - \l_1 \l_2^6 \big) + \mathcal{O}(\l^9) \big]}  \non \\ 
    &\times \bigg[1 +\frac{ \pi^2}{6} \big(-\lambda_1^2 + 6 \lambda_1 \lambda_2\big) + \frac{ \pi^4}{120} \big(\l_1^4 - 10 \lambda_1^3 \lambda_2 - 20 \lambda_1 \lambda_2^3 \big) \non \\ &
    +\frac{ \pi^6}{5040} \big(- \lambda_1^6 + 14 \l_1^5 \l_2
     +700 \lambda_1^3 \lambda_2^3
    + 630 \lambda_1^2 \lambda_2^4 + 42 \lambda_1 \lambda_2^5 \big) \non \\ & + \frac{ \pi^8}{362880} \big( \lambda_1^8 - 18 \lambda_1^7 \lambda_2 - 7728 \lambda_1^5 \lambda_2^3 - 56700
    \lambda_1^4 \lambda_2^4 - 68292 \lambda_1^3 \lambda_2^5  \non \\ & \qquad \quad - 7560 \lambda_1^2 \lambda_2^6 - 72 \lambda_1 \lambda_2^7 \big) ~ +\mathcal{O}(\l^{10})\bigg] 
\end{align}

We stress that this way of rewriting the result really makes sense only if the framing dependence keeps factorizing also at higher loops. This is not a priori guaranteed because of the presence of possible contributions to the framing  coming from vertex--type diagrams, for which an exponentiation theorem does not exist yet.

\end{document}